\documentclass[prd,aps,twocolumn]{revtex4}

\def \d {{\rm d}}
\def \im {{\rm i}} 
\def \boldk {\mbox{\boldmath$k$}} 
\def \boldl {\mbox{\boldmath$l$}} 
\def \boldm {\mbox{\boldmath$m$}} 
 
\newcommand{\scri}{{\mathcal{I}}}

\begin{document}

\title{Photon rockets in (anti-)de~Sitter universe}

\author{J. Podolsk\'y}

\affiliation{Institute of Theoretical Physics, Faculty of Mathematics and Physics, Charles University in Prague, \\
V Hole\v{s}ovi\v{c}k\'{a}ch 2, 180 00 Prague 8, Czech Republic}
\date{\today}

\begin{abstract}
A class of exact solutions of Einstein's equations is presented which describes accelerating photon rockets in de~Sitter and anti-de~Sitter universe. These are particular members of the Robinson--Trautman family of axially symmetric spacetimes with pure radiation.  In particular, generalizations of  (type~D) Kinnersley's rockets and (type~II) Bonnor's rockets to the case of a non-vanishing cosmological constant are given. Some of the main physical properties of these solutions are investigated, and their relation to the C-metric solution which describes uniformly accelerated black holes is also given.
\end{abstract}

\pacs{04.20.Jb, 04.30.Nk}
\maketitle
\narrowtext

\section{Introduction}
\label{intro}

Within the large and important family of Robinson--Trautman spacetimes \cite{RobTra60,RobTra62,Stephanibook} there exist solutions which describe accelerated sources. The best-known of these is the \hbox{C-metric}. It is a particular type~D vacuum solution representing a pair of black holes which uniformly accelerate under the influence of cosmic strings or struts located along the axis  of symmetry~\cite{KinWal70}. Interestingly, there also exist Robinson--Trautman solutions with aligned pure radiation and a regular axis which can be physically interpreted as ``photon rockets''. The first solution of this type was introduced and studied by Kinnersley \cite{Kin69} in 1969. Such a spacetime, that is also of type~D, describes a localized object which emits pure radiation and accelerates due to the net back reaction along a straight line corresponding to the axis. It may thus serve as a simple exact model of a kind of rocket that is propelled by an anisotropic emission of photons. Later, Bonnor \cite{Bon96} presented a generalization of the Kinnersley rocket. His extension also belongs to the family of axially symmetric pure radiation Robinson--Trautman spacetimes, but it is of algebraic type~II. In this case the angular distribution of photons may have much more complicated character that can be prescribed almost arbitrarily.

These solutions have attracted considerable attention after Bonnor's rather surprising discovery that, with respect to the flat background frame, there is no energy loss corresponding to gravitational waves emitted by the accelerating Kinnersley rocket \cite{Bon94}. This absence of gravitational radiation has been subsequently confirmed and clarified by Damour \cite{Dam95} in the context of the associated  problem using the post-Minkowskian perturbation formalism. He argued that, although the Kinnersley rocket emits null fluid anisotropically, it does not produce (linearized) gravitational waves because the matter emission in it has no quadrupole moment. In fact, the radiation is the sum of waves generated by the point-like rocket and of those generated by the distribution of the photon fluid. These two distinct contributions cancel each other. Further studies have been undertaken by Dain, Moreschi and Gleiser \cite{DaiMorGle96} who employed general properties of exact Robinson--Trautman geometries to distinguish, in particular, gravitation and matter-energy radiation. By transforming to Bondi--Sachs coordinates, von~der~G\"onna and Kramer have demonstrated in \cite{vonGonKram98}  that the Kinnersley family of type~D is the only axisymmetric and asymptotically flat Robinson--Trautman solution with pure radiation that does not contain gravitational radiation. This result has been subsequently extended by Cornish \cite{Cor00} who proved that the class of pure radiation Robinson--Trautman metrics for which the news function vanishes is the same as the class of all (not necessarily axially symmetric) Kinnersley photon rocket metrics. Some other aspects of the metric representing a Kinnersley photon rocket have been analyzed in \cite{Car00}.

In all these cases it was assumed that the cosmological constant $\Lambda$ vanishes. It is a well-known fact \cite{Stephanibook} that the cosmological constant can easily be included into the Robinson--Trautman family of solutions. While the corresponding C-metrics with $\Lambda$ have already been thoroughly studied, see e.g. \cite{PodGrif01,Pod02,KrtPod03,PodOrtKrt03,DiaLem03a,DiaLem03b,Krt05}, it seems that the possibility to generalize the Kinnersley and Bonnor rockets to the corresponding spacetimes with a cosmological constant has remained unnoticed. It is the purpose of this work to present such solutions (in Sec.~\ref{Krocketsection} and Sec.~\ref{Brocketsection}, respectively) and to investigate their main properties (Sec.~\ref{propsection}).

\section{Robinson--Trautman space-times with pure radiation}
\label{RTsectionpurerad}

The Robinson--Trautman family of space-times is defined geometrically by the property that it admits a geodesic, shear-free, twist-free but expanding null congruence, denoted here as~${\boldk}$. As shown in \cite{RobTra60,RobTra62,Stephanibook}, a general metric of vacuum or pure radiation spacetime with such properties can be written in the form
 \begin{equation} 
 \d s^2 =2\,{r^2\over P^2}\,\d\zeta\,\d\bar\zeta -2\,\d u\,\d r  -2\,H\,\d u^2\,, 
 \label{RTmetric} 
 \end{equation}
 in which
 \begin{equation} 
 2H = \Delta\log P -2r(\log P)_{,u} -{2m\over r} -{\Lambda\over3}r^2\,, 
 \label{RTHfunction} 
 \end{equation} 
 where \ ${\Delta\equiv 2P^2\partial_{\zeta}\partial_{\bar\zeta}\>}$ and $\Lambda$ is the cosmological constant. This metric contains two functions, \ ${P=P(\zeta,\bar\zeta,u)\>}$ and \ $m=m(u)\,$. The coordinates employed in the metric (\ref{RTmetric}) are adapted to the assumed geometry. Specifically, $r$ is an affine parameter along the rays of the principal null congruence generated by ${\boldk=\partial_r}$, $u$~is a retarded time coordinate, and $\zeta$~is a complex spatial ``stereographic-type'' coordinate. The Gaussian curvature of the 2-surfaces spanned by $\zeta$, on which $u$ is any constant and ${r=1}$, is given by 
\begin{equation}
K(\zeta,\bar\zeta,u)\equiv\Delta\log P\,.
\label{RTGausscurvature}
\end{equation}
For a general fixed value of $r$, the Gaussian curvature of these 2-spaces is ${K(\zeta,\bar\zeta,u)/r^2}$ so that, as ${r\to\infty}$, they (locally) become planar.

Using the null tetrad ${\boldk=\partial_r}$, ${\boldl=\partial_u-H\partial_r}$, ${\boldm=(P/r)\,\partial_\zeta}$ the non-zero components of the curvature tensor for the metric (\ref{RTmetric}) are
 \begin{eqnarray}
 && \Psi_2=-{m\over r^3}\,, \qquad  \Psi_3=-{P\over2r^2}\,K_{,\bar\zeta}\,, \nonumber\\
 && \Psi_4= {1\over 2r^2}\left(P^2K_{,\bar\zeta}\right)_{,\bar\zeta}-{1\over r}\left[ P^2\big(\log P\big)_{,u\bar\zeta} \right]_{,\bar\zeta}\,,\label{RTPsis}\\
 && \Phi_{22}={1\over{4r^2}}[\Delta\Delta\log P+12m(\log P)_{,u} -4m_{,u}]\,.\nonumber
 \end{eqnarray} 
In the present work we will consider Robinson--Trautman space-times which are of algebraic types D or~II. For these, it is required that ${m\ne0}$, and there is a scalar polynomial curvature singularity at ${r=0}$ since ${\Psi_2}$ diverges. The space-time is of type~D if \ ${3\Psi_2\Psi_4=2\Psi_3^{\,2}}$.

The conformal infinity~$\scri$ is given by ${r=\infty}$ where, as can be seen from (\ref{RTPsis}), the space-times become conformally flat and vacuum, i.e. asymptotically Minkowski, de~Sitter or anti-de~Sitter. Indeed, introducing an inverse radial coordinate ${\,l=r^{-1}\,}$ and the conformal factor ${\,\Omega=l\,}$, the Robinson-Trautman metric (\ref{RTmetric}), (\ref{RTHfunction})  becomes
\begin{eqnarray}
&& \Omega^2 \d s^2=\Big(\,\frac{\Lambda}{3}+2l(\log P)_{,u} -l^2 \Delta\log P+ 2ml^3\Big) \d u^2  \nonumber\\
&& \hskip6pc  +2\,\d u\,\d l + 2P^{-2}\d\zeta\,d\bar\zeta\,. \label{confmetricRT}
\end{eqnarray}
Obviously, for smooth $P(\zeta,\bar\zeta,u)$, the conformal infinity $\scri$, which is located at ${\Omega=0}$, is regular because the metric (\ref{confmetricRT}) is regular at ${l=\Omega=0}$. In addition, $\scri$ is null, spacelike or timelike according to the sign of the cosmological constant ${\Lambda=0}$, ${\Lambda>0}$ or ${\Lambda<0}$, respectively.

The non-vanishing Ricci tensor component  $\Phi_{22}$ in (\ref{RTPsis}) corresponds to the presence of an aligned pure radiation field (that is flow of matter of zero rest-mass, emitted from the source at ${r=0}$) with an energy-momentum tensor of the form ${T_{\mu\nu}=\rho\, k_\mu k_\nu\,}$, where ${\boldk=\partial_r}$. The radiation density is given by 
 \begin{equation}
 \rho=\frac{n^2(\zeta,\bar\zeta,u)}{r^2}\,,
 \label{pureradrho}
 \end{equation}
 and the function $n^2$ is determined by the equation
 \begin{equation}
 \Delta\Delta(\log P)+12m(\log P)_{,u}-4m_{,u}=16\pi\, n^2\,,
 \label{RTequationpurerad}
 \end{equation}
while the function $m(u)$ is arbitrary.

As can be seen from expressions (\ref{RTPsis}), there are no type~N or conformally flat solutions of this type. Indeed, the conditions that ${\Psi_2}$ and ${\Psi_3}$ vanish imply that ${m=0}$ and ${\Delta\log P=K(u)}$ which, in turn, imply ${\Phi_{22}=0}$, that is a vacuum space-time with ${\,n=0}$.

\section{The Kinnersley rockets with ${\Lambda}$}
\label{Krocketsection}

The simplest Robinson--Trautman type~D spacetimes with pure radiation are given by
 \begin{equation}
  P=A(u)+B(u)\, \zeta+\bar B(u)\,\bar\zeta+C(u)\,\zeta\bar\zeta\,,
 \label{generalKinnersley}
 \end{equation}
where ${A,B,C}$ are arbitrary functions of $u$, such that 
 \begin{equation}
 K=2(AC-B\bar B)\,.
 \label{KgeneralKinnersley}
 \end{equation} 
For ${K>0}$, these solutions are known as the Kinnersley rockets (see \cite{Stephanibook,Cor00}). They represent the gravitational field of an arbitrarily moving object located at ${r=0}$. 

Here we restrict our attention to the axially symmetric cases with ${\,B=0\,}$ and ${K=+1}$. These correspond to photon rockets moving along a single axis (of symmetry) ${\zeta=0}$. For convenience, it is useful to perform the transformation ${\,\zeta=\exp(\sqrt{2}\,\tilde\zeta)\,}$ and to introduce a new arbitrary ``acceleration'' function $\alpha(u)$ by identification ${\,A(u)\equiv\frac{1}{\sqrt{2}}\exp(\int\!\alpha(u)\d u)\,}$, ${\,C(u)\equiv\frac{1}{\sqrt{2}}\exp(-\int\!\alpha(u)\d u)\,}$. Dropping the tilde over $\zeta$, we thus obtain the Robinson--Trautman family of solutions  determined by the function
  \begin{equation}
  P= \cosh\Big({\textstyle \int\!\alpha(u)\d u - \frac{1}{\sqrt2}(\zeta+\bar\zeta)}\Big),
  \label{specialProcket}
  \end{equation}
for which ${\Psi_3=0=\Psi_4}$. Introducing angular coordinates ${\theta,\phi}$ by
 \begin{equation}
 {\textstyle \cos\theta= \tanh\Big(\int\!\alpha(u)\d u - \frac{1}{\sqrt2}(\zeta+\bar\zeta)\Big), \quad\phi=\frac{-\im}{\sqrt2}(\zeta-\bar\zeta)},
 \label{specialtransfProcket}
 \end{equation}
then ${1/P=\sin\theta}$, ${\Delta\log P=1}$, and ${(\log P)_{,u}=\alpha\cos\theta\,}$. The line element (\ref{RTmetric}), (\ref{RTHfunction}) thus takes the explicit form
 \begin{eqnarray}
 &&\hskip-0.98pc\d s^2 =-\bigg(1-{2\,m\over r} -{\Lambda\over3}r^2-2\alpha\, r\cos\theta -\alpha^2\,r^2\sin^2\theta\bigg)\d u^2  \nonumber\\
 &&\hskip4pc  -\,2\,\d u\,\d r+\>2\alpha\,r^2\sin\theta\,\d u\,\d\theta  \label{Kinnersrocketmetric}\\ 
 &&\hskip4pc  +\,r^2(\d\theta^2+\sin^2\theta\,\d\phi^2)\,,\nonumber
  \end{eqnarray}
 in which $m(u)$ and $\alpha(u)$ are  arbitrary functions of the null coordinate~$u$.

For a pure radiation field ${\,T_{\mu\nu}=\rho\,k_\mu k_\nu\,}$, where $\rho$ is given by (\ref{pureradrho}),  the field equation (\ref{RTequationpurerad}) yields
 \begin{equation}
 3\,\alpha\, m\,\cos\theta-m_{,u}=4\pi\, n^2(\theta,u)\,.
 \label{RTequationKinnersley}
 \end{equation}
 For any given functions $\alpha(u)$ and $m(u)$, the radiation field profile ${n^2(\theta,u)}$ is thus fully determined. Of course, for this to be real for all~$\theta$, it is necessary that ${m_{,u}<-3\,|\alpha\,m|}$, which implies that $m(u)$ must be strictly decreasing.

If ${\alpha=0}$, the metric (\ref{Kinnersrocketmetric}) reduces to 
 \begin{eqnarray}
 &&\d s^2 =-\left(1-{2m(u)\over r}-{\Lambda\over3}r^2\right)\d u^2 \nonumber\\
 &&\hskip4pc  -2\,\d u\,\d r+r^2 (\d\theta^2+\sin^2\theta\d\phi^2)\,,\qquad
 \label{VaidyadeSitter}
 \end{eqnarray}
 which can be recognized as the Vaidya--(anti-)de~Sitter spacetime.  When the cosmological constant vanishes, this reduces to the standard radiating Vaidya metric. It describes a spherically symmetric fixed source with a varying mass determined by $m(u)$ for which the radiation field is given by ${n^2(u)=-\frac{1}{4\pi}\,m_{,u}}$. It is therefore generally assumed that $m$ is a positive and non-increasing function.

When ${\Lambda=0}$, the solution (\ref{Kinnersrocketmetric}) is exactly the particular Kinnersley photon rocket, see \cite{Kin69,Bon94}. It represents a singular source at ${r=0}$, of varying mass determined by $m(u)$, which emits pure radiation and accelerates in Minkowski space along a straight line (which is the axis of symmetry) due to the corresponding net back reaction described by (\ref{RTequationKinnersley}). This thus serves as a simple exact model of a kind of rocket that is propelled by the anisotropic emission of photons.

This interpretation remains valid even in the cases ${\Lambda\not=0}$. Indeed, in the weak field limit in which ${m=0}$, the above spacetime (\ref{Kinnersrocketmetric}) is a conformally flat vacuum solution, namely Minkowski, de~Sitter or anti-de~Sitter space according to the sign of the cosmological constant~$\Lambda$. The function $\alpha(u)$ then determines the acceleration of a test particle located at the origin ${r=0}$ which moves along the timelike trajectory \ ${x^\mu=(u(\tau), r=0, \theta_0, \phi_0)}$, \ where $\tau$ is its proper time, and $\theta_0$ and $\phi_0$ are constants. Normalization requires that \ ${u(\tau)=\tau}$, \ so that the four-velocity is \ ${u^\mu=(1,0,0,0)}$. \ For the corresponding four-acceleration \ ${a^\mu=\Gamma^\mu_{\ uu}}$ \ it follows that \ ${u_\mu a^\mu=0}$ \ and \ ${a_\mu a^\mu=\alpha^2(u)}$. \ This proves that the {\em instantaneous acceleration}\index{acceleration} of the test particle is given by ${\alpha(u)\equiv\alpha(\tau)}$. And, since the function $\alpha(u)$ in the metric (\ref{Kinnersrocketmetric}) can be chosen arbitrarily, it is possible to prescribe any specific acceleration for the source at ${r=0}$. In particular, it is possible to construct a rocket which has a constant acceleration~$\alpha$. 

Further justification of such an interpretation will be given below in Sec.~\ref{propsection}.

\section{The Bonnor rockets with ${\Lambda}$}
\label{Brocketsection}

There exist generalizations of the above Kinnersley photon rockets in the family of type II Robinson--Trautman spacetimes. Restricting attention to axially symmetric pure radiation solutions, it is convenient to introduce the polar-type coordinates $x,\phi$ such that
 \begin{equation}
 \zeta= {1\over\sqrt2} \left( -\int{\d x\over G(x,u)} + \int\!\alpha(u)\d u  +\im\, \phi \right)\,,\label{transRTtoBonnormetric}\\
 \end{equation}
where $G(x,u)$ is an arbitrary function of $x$ and $u$, while $\alpha(u)$ is any function of $u$. With the identification
 \begin{equation}
  P(\zeta,\bar\zeta,u)= G^{-1/2}\Big(x(\zeta,\bar\zeta,u),u\Big)\,,\label{exprPforRTtoBonnormetric}\\
 \end{equation}
in which the function $x(\zeta,\bar\zeta,u)$ is obtained by inverting (\ref{transRTtoBonnormetric}) as
 \begin{equation}
 \int\!{\d x\over G(x,u)} = {\textstyle   \int\!\alpha(u)\d u  -\frac{1}{\sqrt2}(\zeta+\bar\zeta)}\,,
\label{gentransfaceler}
\end{equation}
the standard Robinson--Trautman metric (\ref{RTmetric}), (\ref{RTHfunction}) takes the form
 \begin{eqnarray}
 &&\hskip-1.01pc \d s^2 =\!-\!\left(\!-\frac{1}{2}G_{,xx}-\frac{2m}{r}
 -\frac{\Lambda}{3}r^2-r\,(b\,G)_{,x}-b^2\,G\,r^2\right)\!\d u^2  \nonumber\\
 && \hskip1.2pc  -2\,\d u\,\d r +\> 2\,b\, r^2\,\d u\, \d x  
 +\,r^2\,\Big(\,{\d x^2\over G} +G\,\d\phi^2\Big),
 \label{Bonnmetricexpl}
 \end{eqnarray}
where
 $$ b(x,u)=-\alpha(u)-\int\!\frac{G_{,u}(x,u)}{G^2(x,u)}\,\d x\,. $$

In particular, the Kinnersley rocket (which is of Petrov type~D) presented in Sec.~\ref{Krocketsection} is obtained by taking
\begin{equation}
 G(x,u)=1-x^2\,,
 \label{KinrocketInx}
 \end{equation}
which implies that ${\,b(u)=-\alpha(u)}$. Then, introducing an angular coordinate by putting ${x=\cos\theta}$, the metric (\ref{Kinnersrocketmetric}) is immediately recovered, and the general expressions (\ref{gentransfaceler}) and (\ref{exprPforRTtoBonnormetric}) reduce exactly to (\ref{specialtransfProcket}) and (\ref{specialProcket}) respectively.

The above generalization of the Kinnersley rocket was first presented by Bonnor \cite{Bon96} for the case when ${\Lambda=0}$. Such Bonnor's photon rockets, which also accelerate due to the anisotropic emission of null fluid, are members of the family of axially symmetric Robinson--Trautman type~II solutions (\ref{Bonnmetricexpl}) in which the function $G$ has the form
 \begin{equation}
 G(x,u)=(1-x^2)\Big[1+(1-x^2)\,h(x,u) \Big]\,,
 \label{Bonnorrocket}
 \end{equation}
 where $h(x,u)$ is an arbitrary smooth bounded function (greater than $-1$). Since $G$ must be positive, it is again appropriate to put ${x=\cos\theta}$, and the metric can be seen to be regular on the axis ${x=\pm1}$, which corresponds to ${\theta=0,\pi}$. The function $b$ in~(\ref{Bonnmetricexpl}) then reads
 $$ b(x,u)=-\alpha(u)-\int\!\frac{h_{,u}(x,u)}{[1+(1-x^2)\,h(x,u)]^2}\,\d x\,, $$
 and the pure radiation field, see~(\ref{pureradrho}) and (\ref{RTequationpurerad}), is 
 \begin{equation}
 4\pi\, n^2(x,u)= {\textstyle -\frac{1}{8}(GG_{,xxx})_{,x}+\frac{3}{2}m(b\,G)_{,x}-m_{,u}}\,,
 \label{RTequationBonnor}
 \end{equation}
 which generalizes (\ref{RTequationKinnersley}). The angular distribution of photons thus now has a more complicated character that can be prescribed almost arbitrarily by a specific choice of the smooth function~$h(x,u)$.

\section{Some physical properties}
\label{propsection}

\subsection{Trajectories of accelerated rockets in Minkowski universe}
\label{trajMink} 

It is well-known that, in a weak field limit ${m\to0}$, the Kinnersley solution (\ref{Kinnersrocketmetric}) with ${\Lambda=0}$ reduces to the flat space metric whose origin of coordinates ${r=0}$ is accelerating. An explicit transformation to standard Minkowski coordinates has been described  already by Newman and Unti \cite{NewUnt63}, and subsequently employed for a constant acceleration $\alpha$ in~\cite{KinWal70}, and for a general function $\alpha(u)$ in \cite{Bon94,Bon96} and elsewhere.  It has the form
 \begin{eqnarray}
 &&\hskip-0.7pc T   = \,t(u)\! + r\Big[\!-\cos\theta  \sinh{\!\big({\textstyle   \int\!\alpha(u)\d u }\big)}\!+\cosh{\!\big({\textstyle   \int\!\alpha(u)\d u }\big)} \Big]\!, \nonumber\\
 &&\hskip-0.7pc X \!= x(u) \!+  r\Big[\!-\cos\theta  \cosh{\!\big({\textstyle   \int\!\alpha(u)\d u }\big)}\!+\sinh{\!\big({\textstyle   \int\!\alpha(u)\d u }\big)}\Big]\!, \nonumber\\
 &&\hskip-0.7pc Y   = r\sin\theta\cos\phi\,, \nonumber\\
 &&\hskip-0.7pc Z   = r\sin\theta\sin\phi\,, \label{accelbackMINK}
  \end{eqnarray}
where the functions $t(u)$ and $x(u)$ are given by
\begin{equation}
 \dot t=\cosh{\left({\textstyle   \int\!\alpha(u)\d u }\right)}\,,\quad
 \dot x=\sinh{\left({\textstyle   \int\!\alpha(u)\d u }\right)}  \,,
 \label{accelerparameters}
 \end{equation}
with the dot denoting a differentiation with respect to~$u$. For vanishing acceleration, ${\alpha=0}$, the transformation reduces to the standard relation between spherical and cartesian coordinates in flat space. Obviously, the (test) rocket located at ${r=0}$ moves along the trajectory  
\begin{eqnarray}
 && T   = \,t(u)\,, \nonumber\\
 && X \!= x(u)\,, \label{trajectaccelbackMINK}\\
 && Y = 0 =Z\,, \nonumber
  \end{eqnarray} 
 which describes a {\em generally accelerated motion along the axis of symmetry} (with a normalized four-velocity). In particular, for a {\em uniform} acceleration ${\alpha=\,}$const. the relations~(\ref{accelbackMINK}) simplify to
 \begin{eqnarray}
 && T   = \left(\alpha^{-1} - r\cos\theta \right) \sinh(\alpha\,u) + r\cosh{(\alpha\, u)}\,, \nonumber\\
 && X \!= \left(\alpha^{-1} - r\cos\theta \right) \cosh(\alpha\, u) + r\sinh{(\alpha\, u)}\,, \nonumber\\
 && Y   = r\sin\theta\cos\phi\,, \label{uniformaccelbackMINK}\\
 && Z   = r\sin\theta\sin\phi\,, \nonumber
  \end{eqnarray}
and the rocket at ${r=0}$ moves along
\begin{eqnarray}
 && T   = \alpha^{-1} \sinh(\alpha\, u)\,, \nonumber\\
 && X \!= \alpha^{-1} \cosh(\alpha\, u)\,, \label{uniformtrajectaccelbackMINK}\\
 && Y = 0 =Z\,, \nonumber
\end{eqnarray} 
which is the usual trajectory.

\subsection{Trajectories of uniformly accelerated rockets in (anti-)de~Sitter universe}
\label{trajunifadS} 

A similar interpretation can also be given when the cosmological constant $\Lambda$ is nonvanishing. In such a case, the rocket with a negligible mass $m$ can be considered as a test particle moving in the de~Sitter (${\Lambda>0}$) or anti-de~Sitter (${\Lambda<0}$) background universe. These maximally symmetric, conformally flat spacetimes of constant curvature ${R=4\Lambda}$ can be represented as a four-dimensional hyperboloid 
 \begin{equation}
 -Z_0^2 +Z_1^2 +Z_2^2 +Z_3^2 +\epsilon\,Z_4^2 =\frac{3}{\Lambda} \,,
 \label{hyperboloid} 
 \end{equation}
embedded in a five-dimensional flat space described by the metric
\begin{equation}
 \d s^2= -\d{Z_0}^2 +\d{Z_1}^2 +\d{Z_2}^2 +\d{Z_3}^2 +\epsilon\,\d{Z_4}^2\,, 
 \label{5DimMink} 
 \end{equation} 
 where ${\,\epsilon\equiv\hbox{sign}\,\Lambda\,}$. For ${m=0}$ and a {\em uniform acceleration} ${\,\alpha=\,}$const. such that ${\alpha^2+{\textstyle\frac{1}{3}}\Lambda>0}$, the coordinates employed in the solution (\ref{Kinnersrocketmetric}) are given by the following parametrization of the (anti-)de~Sitter space,   
 \begin{eqnarray}
 && Z_0  = \frac{1 - \alpha\, r\cos\theta}{\sqrt{\alpha^2+{\textstyle\frac{1}{3}}\Lambda}}\, \sinh\big(\sqrt{\alpha^2+{\textstyle\frac{1}{3}}\Lambda}\, u \big) \nonumber\\
   &&\hskip5.4pc  +\, r\,\cosh{\big(\sqrt{\alpha^2+{\textstyle\frac{1}{3}}\Lambda}\, u\big)}, \nonumber\\
 && Z_1  = \frac{1 - \alpha\, r\cos\theta}{\sqrt{\alpha^2+{\textstyle\frac{1}{3}}\Lambda}}\, 
\cosh\big(\sqrt{\alpha^2+{\textstyle\frac{1}{3}}\Lambda}\, u \big) \nonumber\\
   &&\hskip5.4pc + \,r\,\sinh{\big(\sqrt{\alpha^2+{\textstyle\frac{1}{3}}\Lambda}\, u\big)}, \nonumber\\
 && Z_2  = r\sin\theta\cos\phi\,, \label{accelbackADS}\\
 && Z_3  = r\sin\theta\sin\phi\,, \nonumber\\
 && Z_4  = \frac{ \alpha +{{\textstyle\frac{1}{3}}\Lambda}\, r\cos\theta  }
{\sqrt{{\textstyle\frac{1}{3}}|\Lambda|}\,\sqrt{\alpha^2+{\textstyle\frac{1}{3}}\Lambda}}\,. \nonumber
  \end{eqnarray} 
These relations generalize expressions (\ref{uniformaccelbackMINK}), and reduce to them when ${\Lambda=0}$ (ignoring the fixed value of $Z_4$). The rocket located at ${r=0}$ obviously moves along the trajectory   
\begin{eqnarray}
 && Z_0   = \frac{1}{\sqrt{\alpha^2+{\textstyle\frac{1}{3}}\Lambda}}\, \sinh\big(\sqrt{\alpha^2+{\textstyle\frac{1}{3}}\Lambda}\, u\big),  \nonumber\\
 && Z_1   = \frac{1}{\sqrt{\alpha^2+{\textstyle\frac{1}{3}}\Lambda}}\, \cosh\big(\sqrt{\alpha^2+{\textstyle\frac{1}{3}}\Lambda}\, u\big), 
\nonumber\\
 && Z_2 = 0 =Z_3\,, \label{trajectaccelbackADS}\\
 && Z_4  = \frac{\alpha}{\sqrt{{\textstyle\frac{1}{3}}|\Lambda|}\,\sqrt{\alpha^2+{\textstyle\frac{1}{3}}\Lambda}}\,, \nonumber
  \end{eqnarray}
in the de~Sitter or anti-de~Sitter universe. Such a trajectory describes uniformly accelerated motion along the axis of symmetry, cf. \cite{DiaLem03b,PodGrif06}.

In the case of a test particle moving with small constant acceleration $\alpha$ in the anti-de~Sitter universe, for which ${\alpha^2+{\textstyle\frac{1}{3}}\Lambda<0}$, the metric (\ref{Kinnersrocketmetric}) corresponds to the parametrization
\newpage
 \begin{eqnarray}
 && Z_0  = \frac{1 - \alpha\, r\cos\theta}{\sqrt{-(\alpha^2+{\textstyle\frac{1}{3}}\Lambda)}}\, \sin\big(\sqrt{-(\alpha^2+{\textstyle\frac{1}{3}}\Lambda)}\, u\big) \nonumber\\
   &&\hskip6.2pc  +\, r\,\cos{\big(\sqrt{-(\alpha^2+{\textstyle\frac{1}{3}}\Lambda)}\, u\big)}, \nonumber\\
 && Z_1  = \frac{ \alpha +{{\textstyle\frac{1}{3}}\Lambda}\, r\cos\theta  }
{\sqrt{{\textstyle\frac{1}{3}}|\Lambda|}\,\sqrt{-(\alpha^2+{\textstyle\frac{1}{3}}\Lambda})}\,, \nonumber\\
 && Z_2  = r\sin\theta\cos\phi\,, \label{accelbacksmallADS}\\
 && Z_3  = r\sin\theta\sin\phi\,, \nonumber\\
 && Z_4  = \frac{1 - \alpha\, r\cos\theta}{\sqrt{-(\alpha^2+{\textstyle\frac{1}{3}}\Lambda)}}\, \cos\big(\sqrt{-(\alpha^2+{\textstyle\frac{1}{3}}\Lambda)}\, u\big) \nonumber\\
   &&\hskip6.2pc  -\, r\,\sin{\big(\sqrt{-(\alpha^2+{\textstyle\frac{1}{3}}\Lambda)}\, u\big)}. \nonumber
  \end{eqnarray}
The rocket thus moves along the uniformly accelerated trajectory   
\begin{eqnarray}
 && Z_0   = \frac{1}{\sqrt{-(\alpha^2+{\textstyle\frac{1}{3}}\Lambda)}}\, \sin\big(\sqrt{-(\alpha^2+{\textstyle\frac{1}{3}}\Lambda)}\, u\big),  \nonumber\\
 && Z_4   = \frac{1}{\sqrt{-(\alpha^2+{\textstyle\frac{1}{3}}\Lambda)}}\, \cos\big(\sqrt{-(\alpha^2+{\textstyle\frac{1}{3}}\Lambda)}\, u\big), \nonumber\\
&& Z_1  = \frac{\alpha}{\sqrt{{\textstyle\frac{1}{3}}|\Lambda|}\,\sqrt{-(\alpha^2+{\textstyle\frac{1}{3}}\Lambda)}}\,, \label{trajectaccelbacksmallADS}\\ 
&& Z_2 = 0 =Z_3\,, \nonumber
\end{eqnarray}
in the anti-de~Sitter universe. In this case, it can never reach the conformal infinity located at ${Z_1=\infty}$, see e.g. \cite{Pod02,DiaLem03a,PodOrtKrt03,Krt05}.

\subsection{Trajectories of generally accelerated rockets in de~Sitter universe}
\label{trajgeneradS}

For the rocket moving with a generally {\em non-uniform acceleration} given by an arbitrary function $\alpha(u)$, the explicit parametrization of the hyperboloid (\ref{hyperboloid}) corresponding to the metric (\ref{Kinnersrocketmetric}) is more complicated. 

However, in the case ${\Lambda>0}$ it is possible first to consider the relations
\begin{eqnarray}
 && Z_0  = z_0(u) + r\,\frac{1}{P(u,\vartheta)}\,, \nonumber\\
 && Z_1  = z_1(u) + r\,\frac{\cos\vartheta}{P(u,\vartheta)}\,, \nonumber\\
 && Z_2  = r\,\frac{\sqrt{\sin^2\vartheta-Q^2(u,\vartheta)}}{P(u,\vartheta)}\,\cos\phi\,, \label{genaccelbackADS}\\
 && Z_3  = r\,\frac{\sqrt{\sin^2\vartheta-Q^2(u,\vartheta)}}{P(u,\vartheta)}\,\sin\phi\,. \nonumber\\
 && Z_4  = z_4(u) + r\,\frac{Q(u,\vartheta)}{P(u,\vartheta)}\,, \nonumber
  \end{eqnarray} 
 where the functions $z_0(u)$, $z_1(u)$ and $z_4(u)$ satisfy the constraint
 \begin{equation}
 -z_0^2 +z_1^2 +z_4^2 =\frac{3}{\Lambda} \,,
 \label{constraint5}
 \end{equation}
 but otherwise are completely arbitrary. The functions $P$,~$Q$ are given by
 \begin{eqnarray}
 && P(u,\vartheta)=z_4(u)\,\big[\,\dot f(u)-\dot g(u)\cos\vartheta\,\big],  \nonumber\\
 && Q(u,\vartheta)=f(u)-g(u)\cos\vartheta\,, \label{skoroparametr}
  \end{eqnarray}
in which
 \begin{equation}
 f(u)= \frac{z_0(u)}{z_4(u)} \,,\qquad g(u)= \frac{z_1(u)}{z_4(u)}\,.
 \label{deffag}
 \end{equation}
This obviously corresponds to general motion of a test particle located at ${r=0}$ on the hyperboloid (\ref{hyperboloid}) such that ${Z_2 = 0 =Z_3}$. 
Moreover, (\ref{genaccelbackADS}) is a regular parametrization of the de~Sitter universe by the metric 
 \begin{eqnarray}
 &&\hskip-1.3pc\d s^2 =-\bigg(1-2\,r\,(\log P)_{,u} -r^2\,\frac{\sin^2\vartheta}{R}\bigg)\d u^2  \nonumber\\
 &&\hskip1.1pc  -\,2\,\d u\,\d r+\,2\,r^2\,\frac{z_1-z_0\cos\vartheta}{P\,R}\sin\vartheta\,\d u\,\d\vartheta \nonumber \\  &&\hskip1.1pc +\,r^2\left(\frac{\frac{3}{\Lambda}\sin^2\vartheta}{P^2\,R}\,\d\vartheta^2+\frac{\sin^2\vartheta-Q^2}{P^2}\,\d\phi^2\right)\!,\label{almostKinnersrocketmetric}
  \end{eqnarray} 
where 
\begin{equation}
 R(u,\vartheta)=z_4^2(u)\,[\,\sin^2\vartheta-Q^2(u,\vartheta)\,]\,.
 \label{defR}
 \end{equation}

 Now, the subsequent transformation
\begin{equation}
 \sin\theta=\frac{\sqrt{\sin^2\vartheta-Q^2(u,\vartheta)}}{P(u,\vartheta)}\,,
 \label{finaltransf}
 \end{equation} 
 brings the metric (\ref{almostKinnersrocketmetric}) to the general form (\ref{Kinnersrocketmetric}). In fact, the relations (\ref{genaccelbackADS}) then become
\begin{eqnarray}
 && Z_0  = z_0(u) + r\,W_0(u,\theta)\,, \nonumber\\
 && Z_1  = z_1(u) + r\,W_1(u,\theta)\,, \nonumber\\
 && Z_2  = r\,\sin\theta\,\cos\phi\,, \label{genaccelbackADSW}\\
 && Z_3  = r\,\sin\theta\,\sin\phi\,, \nonumber\\
 && Z_4  = z_4(u) + r\,W_4(u,\theta)\,, \nonumber
  \end{eqnarray} 
but the explicit functions ${W_i(u,\theta)}$ are rather complicated because ${\vartheta(u,\theta)}$ must be obtained by inverting the transformation (\ref{finaltransf}). Interestingly, this leads to a quadratic equation in ${\,\cos\vartheta\,}$ which can be solved as  
\begin{equation}
 \cos\vartheta=\frac{fg+\dot f\dot g\, z_4^2\sin^2\theta+S}{1+g^2+\dot g^2 z_4^2\sin^2\theta}\,,
 \label{quadratcosin}
\end{equation}
where
\begin{equation}
 S^2=1-f^2+g^2-\big[\dot f^2-\dot g^2+(\dot f g-f\dot g)^2\big] z_4^2\sin^2\theta\,.
 \label{quadratcos}
\end{equation} 
A straightforward calculation then shows that
\begin{eqnarray}
  && W_0  = \frac{1}{P}=z_4^{-1}\frac{1+g^2+\dot g^2 z_4^2\sin^2\theta}{\dot f+g(\dot f g-f\dot g)-\dot g\, S}\,, \nonumber\\
  && W_1  = \frac{\cos\vartheta}{P}=z_4^{-1}\frac{fg+\dot f\dot g\, z_4^2\sin^2\theta+S}{\dot f+g(\dot f g-f\dot g)-\dot g\, S}\,, \label{genADSWcompl}\\
  && W_4  = \frac{Q}{P}=z_4^{-1}\frac{f-\dot g(\dot f g-f\dot g)\, z_4^2\sin^2\theta-g\,S}{\dot f+g(\dot f g-f\dot g)-\dot g\, S}\,, \nonumber
  \end{eqnarray} 
which are rather involved expressions.

In particular, special motion of uniform acceleration ${\,\alpha=\,}$const. in the de~Sitter universe is given by the functions 
\begin{eqnarray}
 && z_0(u)   = \frac{1}{\sqrt{\alpha^2+{\textstyle\frac{1}{3}}\Lambda}}\, \sinh\big(\sqrt{\alpha^2+{\textstyle\frac{1}{3}}\Lambda}\, u\big),  \nonumber\\
 && z_1(u)   = \frac{1}{\sqrt{\alpha^2+{\textstyle\frac{1}{3}}\Lambda}}\, \cosh\big(\sqrt{\alpha^2+{\textstyle\frac{1}{3}}\Lambda}\, u\big), 
\label{trajectaccelbackADSagain}\\
 && z_4  = \frac{\alpha}{\sqrt{{\textstyle\frac{1}{3}}\Lambda}\,\sqrt{\alpha^2+{\textstyle\frac{1}{3}}\Lambda}}\,, \nonumber
  \end{eqnarray}
see~(\ref{trajectaccelbackADS}), which satisfy the condition (\ref{constraint5}). Obviously,
\begin{eqnarray}
 && \dot z_0  = \cosh \big(\sqrt{\alpha^2+{\textstyle\frac{1}{3}}\Lambda}\, u\big)\,, \nonumber\\
 && \dot z_1  = \sinh \big(\sqrt{\alpha^2+{\textstyle\frac{1}{3}}\Lambda}\, u\big)\,, \label{paramvelocitspec}\\
 && \dot z_4  = 0\,, \nonumber
  \end{eqnarray}  
so that ${-\dot z_0^2 +\dot z_1^2 +\dot z_4^2 =-1}$, and $u$ is thus the proper time of the rocket. In such a case, the function $S$ given by (\ref{quadratcos}) reduces to 
\begin{equation} 
S=\pm\frac{\sqrt{\alpha^2+{\textstyle\frac{1}{3}}\Lambda}}{\alpha}\,\cos\theta\,.
 \label{specS}
\end{equation} 
With the lower sign, the functions (\ref{genADSWcompl}) simplify considerably and factorize to 
\begin{eqnarray}
  && W_0  = \cosh\beta - \frac{\alpha\, \cos\theta}{\sqrt{\alpha^2+{\textstyle\frac{1}{3}}\Lambda}}\, \sinh\beta\,, \nonumber\\
  && W_1  = \sinh\beta - \frac{\alpha\, \cos\theta}{\sqrt{\alpha^2+{\textstyle\frac{1}{3}}\Lambda}}\, \cosh\beta\,, \label{genADSWcomplspec}\\  
  && W_4  = \frac{\sqrt{{\textstyle\frac{1}{3}}\Lambda}\, \cos\theta}{\sqrt{\alpha^2+{\textstyle\frac{1}{3}}\Lambda}}\,, \nonumber
  \end{eqnarray}  
where ${\,\beta(u)=\sqrt{\alpha^2+{\textstyle\frac{1}{3}}\Lambda}\,\,u \,}$. Inserting these functions into (\ref{genaccelbackADSW}), the simple explicit expressions (\ref{accelbackADS}) are recovered.

Of course, analogous motion of a generally accelerating photon rocket in the anti-de~Sitter universe can be described similarly.

\subsection{Character of null surfaces}
\label{nullcones} 

Using (\ref{finaltransf}), it immediately follows from expressions (\ref{genaccelbackADSW}) and (\ref{genADSWcompl}) that
\begin{eqnarray}
&& -\left[Z_0 - z_0(u)\right]^2+\left[Z_1 - z_1(u)\right]^2+Z_2^2+Z_3^2  \label{null cones}\\
&& \hskip10pc +\left[Z_4 - z_4(u)\right]^2=0\,.
 \nonumber
 \end{eqnarray}
This clearly demonstrates that the surfaces $u=\,$const. form a family of null cones with vertices on the timelike trajectory that represents the origin of the coordinates ${\,r=0}$ where the photon rocket is located. Such future-oriented  null cones extending from these vertices naturally foliate the background de~Sitter space. A similar interpretation applies to the anti-de~Sitter space. 

Recall that $r$ is an affine parameter along the rays tangent to $u=\,$const., which are  generated by the principal null vector field ${\boldk=\partial_r}$, and $\theta$, $\phi$~are the remaining spatial coordinates (the latter corresponding to the orbits of the Killing vector field ${\partial_\phi}$ of these axially symmetric spacetimes).

\subsection{Relation to the C-metric}
\label{Cmetric} 

Within the family of Robinson--Trautman type~D spacetimes there are vacuum solutions called the \hbox{C-metric}. These describe black holes uniformly accelerating in Minkowski, de~Sitter or anti-de~Sitter universe under the influence of cosmic strings or struts located along the axis of symmetry~\cite{KinWal70,PodGrif01,Pod02,KrtPod03,PodOrtKrt03,DiaLem03a,DiaLem03b,Krt05}. Indeed, the transformation (see \cite{KrtPod03}) 
 \begin{eqnarray}
  && r= {1\over \alpha(x+y)}\,,\nonumber\\
  && u= {1\over \alpha} \left( \tau+\int{\d y\over F} \right) ,\label{transRTtoCmetric}\\
  && \zeta= {1\over\sqrt2} \left(  \tau+\int{\d y\over F}-\int{\d x\over G}  +\im\phi \right),\nonumber
 \end{eqnarray} 
 where 
 \begin{eqnarray}
  && G(x) = (1-x^2)(1+2\alpha mx)\,,\nonumber\\
  && F(y) = -{\Lambda\over3\alpha^2}-(1-y^2)(1-2\alpha my)\,,\label{FandG}
 \end{eqnarray} 
 puts the Robinson--Trautman line element (\ref{RTmetric}), (\ref{RTHfunction}) into the familiar form of the \hbox{C-metric}, namely 
\begin{equation} 
\d s^2={1\over\alpha^2(x+y)^2}\left( -F\,\d \tau^2  +{\d y^2\over F} +{\d x^2\over G} +G\,\d\phi^2  \right) .
 \label{Cmetric standard}
\end{equation} 
Here, ${2H=\alpha^2r^2(F+G)}$ and ${P= G^{-1/2}\big(x(\zeta,\bar\zeta,u)\big)}$, 
 where the function $x(\zeta,\bar\zeta,u)$ is obtained from (\ref{transRTtoCmetric}) as ${\int{G(x)^{-1}\d x} ={\textstyle  \alpha\, u -\frac{1}{\sqrt2}(\zeta+\bar\zeta)}}$,
 cf. (\ref{exprPforRTtoBonnormetric}) and (\ref{gentransfaceler}).

An alternative form of the C-metric is obtained by keeping the original Robinson--Trautman coordinates ${r, u}$ and introducing only the coordinates ${x, \phi}$ by the transformation (\ref{transRTtoCmetric}), namely ${\int{G(x)^{-1}\d x-\im\phi } =\alpha\, u -\sqrt2\,\zeta}$. This leads to the line element 
 \begin{eqnarray} 
 && \hskip-1pc \d s^2 =-(2H-\alpha^2r^2\,G)\,\d u^2-2\,\d u\,\d r  - 2\,\alpha\, r^2\,\d u\, \d x \nonumber\\[2pt]
 && \hskip8pc {\displaystyle +\,r^2\,\Big({\d x^2\over G} +G\,\d\phi^2\Big), } 
  \label{KWCmetric} 
 \end{eqnarray} 
 which, for ${\Lambda=0}$, was introduced by Kinnersley and Walker \cite{KinWal70}. 

The solution (\ref{Kinnersrocketmetric}), which represents accelerating Kinnersley photon rockets with $\Lambda$,  {\em is only indirectly} related to the C-metric, which describes uniformly accelerating black holes. The main difference is that the Kinnersley rocket spacetime contains pure (null) radiation whereas the \hbox{C-metric} is a vacuum solution. When ${n=0}$ in the field equation (\ref{RTequationKinnersley}), the spacetime is vacuum and the only possibilities are given by ${\alpha=0}$, $m$ constant, which yields the Schwarzschild--(anti-)de~Sitter spacetime, or ${m=0}$, which corresponds to the Minkowski or \hbox{(anti-)}de Sitter universe. Also, the complete axis given by ${\theta=0, \pi}$ in (\ref{Kinnersrocketmetric}) is regular, without strings or struts, while the \hbox{C-metric} spacetime must always include a topological singularity of some kind on (at least one half of) the axis, which acts as a different ``physical cause'' of black hole acceleration. Nevertheless, these two distinct classes of solutions involving accelerated sources are similar. Both of these type~D spacetimes can be written in the common form (\ref{KWCmetric}). The Kinnersley photon rocket (\ref{Kinnersrocketmetric}) is obtained for \ ${x=\cos\theta}$, \ ${G=1-x^2}$ \ and \ ${2H=1-2\,m(u)\,r^{-1} -{\Lambda\over3}r^2-2\alpha(u)\, r\, x}$.

Finally, we also notice that the metric (\ref{KWCmetric}) is a special case of the type~II metric (\ref{Bonnmetricexpl}) if $G(x)$ is independent of~$u$ and thus ${b(u)=-\alpha(u)}$.

\section{Conclusions}
\label{RTfinalsection}

We presented metrics (\ref{Kinnersrocketmetric}) and (\ref{Bonnmetricexpl}) as generalizations of previously investigated families of  Kinnersley's rockets (of algebraic type~D) and Bonnor's rockets (of type~II) to the case of a non-vanishing cosmological constant~$\Lambda$. These are axially symmetric exact spacetimes of the Robinson--Trautman type that describe photon rockets moving arbitrarily in de~Sitter or anti-de~Sitter universe due to anisotropic emission of pure (null) radiation.

We analyzed the character of such spacetimes, in particular the trajectories of the photon rockets. Treated as test particles, they accelerate in the corresponding background spaces of constant curvature, that is in Minkowski or (anti-)de~Sitter universes, according to the sign of cosmological constant. We gave explicit coordinate parametrizations of the de~Sitter and anti-de~Sitter hyperboloids (\ref{accelbackADS}), (\ref{accelbacksmallADS}) in the case of uniform acceleration, and also for a completely general motion along the axis of symmetry, see (\ref{genaccelbackADSW}), (\ref{genADSWcompl}).

We also elucidated the relation of these pure radiation spacetimes, which describe accelerating photon rockets, to the well-known C-metric family of vacuum solutions, which represent uniformly accelerating black holes.

\section{Acknowledgements}

I am grateful to Jerry Griffiths for his very useful comments. This work was partly supported by the grants GA\v{C}R~202/06/0041, GA\v{C}R~202/08/0187, and by the Czech Centre for Theoretical Astrophysics LC06014.


\begin{thebibliography}{20}

\bibitem{RobTra60}
I. Robinson and A. Trautman, 
Spherical gravitational waves, 
Phys. Rev. Lett. {\bf 4}, 431--432 (1960).

\bibitem{RobTra62}
I. Robinson and A. Trautman, 
Some spherical gravitational waves in general relativity, 
Proc. Roy. Soc. A {\bf 265}, 463--473 (1962). 

\bibitem{Stephanibook}
H. Stephani, D. Kramer, M. MacCallum, C. Hoenselaers and E. Herlt, 
{\em Exact Solutions of Einstein's Field Equations}, 2nd Edition 
(Cambridge University Press, Cambridge, England, 2003).

\bibitem{KinWal70} 
W. Kinnersley and M. Walker, 
Uniformly accelerating charged mass in general relativity, 
Phys. Rev. D {\bf 2}, 1359--1370  (1970).

\bibitem{Kin69} 
W. Kinnersley, 
Field of an arbitrarily accelerating point mass, 
Phys. Rev. {\bf 186}, 1335--1336  (1969).

\bibitem{Bon96}  
W. B. Bonnor, 
Another photon rocket,
Class. Quantum Grav. {\bf 13}, 277--282  (1996). 

\bibitem{Bon94}
W. B. Bonnor, 
The photon rocket,
Class. Quantum Grav. {\bf 11}, 2007--2012  (1994).

\bibitem{Dam95}
T. Damour, 
Photon rockets and gravitational radiation, 
Class. Quantum Grav. {\bf 12}, 725--737  (1995).

\bibitem{DaiMorGle96}
S. Dain, O. M. Moreschi and R. J. Gleiser, 
Photon rockets and the Robinson--Trautman geometries, 
Class. Quantum Grav. {\bf 13}, 1155--1160  (1996).

\bibitem{vonGonKram98}
U. von~der~G\"onna and D. Kramer, 
Pure and gravitational radiation, 
Class. Quantum Grav. {\bf 15}, 215--223 (1998).

\bibitem{Cor00}
F. H. J. Cornish, 
Robinson--Trautman radiating metrics with zero news and photon rockets, 
Class. Quantum Grav. {\bf 17}, 3945--3950  (2000).

\bibitem{Car00}
S. Carlip, 
Aberration and the speed of gravity, 
Phys. Lett. A {\bf 267}, 81--87  (2000).

\bibitem{PodGrif01}
J. Podolsk\'y and J. B. Griffiths, 
Uniformly accelerating black holes in a de~Sitter universe, 
Phys. Rev. D {\bf 63}, 024006  (2001).

\bibitem{Pod02}  
J. Podolsk\'y, 
Accelerating black holes in anti-de Sitter universe,
Czech. J. Phys. {\bf 52}, 1--10 (2002).

\bibitem{KrtPod03}
P. Krtou\v{s} and J. Podolsk\'y, 
Radiation from accelerating black holes in a de Sitter universe, 
Phys. Rev. D {\bf 68}, 024005 (2003).

\bibitem{PodOrtKrt03}
J. Podolsk\'y, M. Ortaggio and  P. Krtou\v{s},
Radiation from accelerated black holes in an anti-de Sitter universe,
Phys. Rev. D {\bf 68}, 124004 (2003).
  
\bibitem{DiaLem03a}
\'O. J. C. Dias and J. P. S. Lemos, 
Pair of accelerated black holes in an anti-de~Sitter background: the AdS $C$ metric, 
Phys. Rev. D {\bf 67}, 064001 (2003).

\bibitem{DiaLem03b}
\'O. J. C. Dias and J. P. S. Lemos, 
Pair of accelerated black holes in a de~Sitter background: the dS $C$ metric,
Phys. Rev. D  {\bf 67}, 084018 (2003).

\bibitem{Krt05}
P. Krtou\v{s}, 
Accelerated black holes in an anti-de Sitter universe,
Phys. Rev. D {\bf 72}, 124019 (2005).

\bibitem{NewUnt63}
E. T. Newman and T. W. J. Unti, 
A class of null flat-space coordinate systems, 
J. Math. Phys. {\bf 4}, 1467--1469 (1963).

\bibitem{PodGrif06}
J. Podolsk\'y and J. B. Griffiths, 
Accelerating Kerr--Newman black holes in (anti-)de Sitter space-time, 
Phys. Rev. D {\bf 73}, 044018 (2006).


\end{thebibliography}
\end{document}